\documentclass[11pt]{article}
\usepackage{hyperref}
\usepackage{amssymb}
\usepackage{graphicx}
\usepackage{amsmath}
\usepackage{authblk}
\usepackage{cite}
\usepackage{physics}
\usepackage[section]{placeins}
\usepackage[a4paper, total={6.8in, 10in}]{geometry}
\numberwithin{equation}{section}
\providecommand{\keywords}[1]
{
  \small	
  \textbf{\textit{Keywords---}} #1
}
\title{$\kappa$-deformed power spectrum and modified Unruh temperature}
\author {Vishnu Rajagopal \thanks{vishnurajagopal.anayath@gmail.com}}
\affil{Deenbandhu Chhotu Ram University of Science and Technology\\Murthal, Sonepat, Haryana - 131039, India}
\date{}
\begin{document}
\maketitle
\begin{abstract}
We study the power spectrum of the uniformly accelerating scalar field, obeying the $\kappa$-deformed Klein-Gordon equation. From this we obtain the $\kappa$-deformed corrections to the Unruh temperature, valid up to first order in the $\kappa$-deformation parameter $a$. We also show that in the small acceleration limit, this expression for the Unruh temperature in $\kappa$-deformed space-time is in exact agreement with the one derived from the $\kappa$-deformed uncertainty relation. Finally, we obtain an upper bound on the deformation parameter $a$.
\end{abstract}
\section{Introduction}

A complete theory of quantum gravity is required to understand the space-time structure at high energy scales. Various approaches such as string theory, loop quantum gravity, asymptotically safe gravity, etc., have been devised in the past and recent decades to formulate a quantum theory of gravity. Almost all of these quantum gravity models have predicted the existence of a minimal length scale beyond which the space-time has no operational meaning. At this juncture the usual notion of the space-time manifold can be replaced with a non-commutative space-time geometry, which incorporates this minimal length scale naturally into it \cite{snyder,connes,doplicher}.

Different types of NC space-times have been constructed and their implications have been studied in detail in the past decades. Moyal space-time is a well explored canonical type NC space-time that has been found to emerge in the low energy limit of string theory \cite{sieberg}. Moyal space-time violates the Lorentz symmetry and its symmetry algebra is defined using the idea of twisted Poincare algebra \cite{chai}, which will reduce to the usual Poincare algebra in the appropriate commutative or low energy limit. The $\kappa$-deformed space-time is a Lie algebraic type NC space-time that has been shown to appear in the low energy limit of loop quantum gravity models \cite{loop}. The symmetry corresponding to the $\kappa$-Minkowski space-time has been realised using the deformed Poincare symmetries such as $\kappa$-Poincare algebra \cite{wjk,majid}. Various NC quantum field theory models compatible with the $\kappa$-Poincare algebra have been constructed and its other implications in the $\kappa$-Minkowski space-time have been investigated in the recent times \cite{gauge0,kreal,mel1,mel3,mel4,mel5}.
   
The vacuum plays a crucial role in developing different quantum field theory models. From the Unruh effect it has been understood that the vacuum of the quantum fields are observer dependent. A uniformly accelerating observer will find the vacuum to be in a thermal bath, whose temperature is known as the Unruh temperature \cite{unruh}. So it is very important to understand the changes in the behaviour of the vacuum due to the quantum gravity effects. Such information about the vacuum can be revealed to some extend by calculating the modifications to the Unruh effect, caused by the quantum gravity effects. 

Recently the Unruh effect has been studied in various NC space-times. In \cite{sachin}, the Unruh effect in Moyal space-time has been studied by calculating the thermal correlation function of the uniformly accelerating twisted scalar field, using the method of Bogoliubov coefficients. Unruh effect in Doplicher-Fredenhagen-Roberts (DFR) space-time (which is a Lorentz invariant extension of Moyal space-time) has been studied by generalising the detector-field interaction method to DFR space-time \cite{vishnu2} and it has been shown that statistics of the distribution function depends upon the number of the extra spatial dimensions introduced by the non-commutativity. The $\kappa$-deformed corrections to the Unruh effect has been studied in \cite{yee,verma1}, by analysing the response of a uniformly accelerating detector coupled to scalar field admitting the $\kappa$-deformed Klein-Gordon equation. The $\kappa$-deformed corrections to Unruh effect using the fermionic field, satisfying the $\kappa$-Dirac equation has also been obtained in \cite{verma2}. In \cite{vishnu1}, the notion of the $\kappa$-deformed oscillator algebra has been exploited to study Unruh effect in the $\kappa$-deformed space-time, by using Bogoliubov coefficients. In \cite{glik}, the $\kappa$-deformations to the thermal spectrum associated with the accelerating observer has been studied using the star product formalism compatible with $\kappa$-Minkowski space-time. 

In this work, we derive the $\kappa$-deformed corrections to the Unruh temperature by generalising the power spectrum method to the $\kappa$-deformed space-time. Rather than using the star product formalism as in \cite{glik}, here we construct the power spectrum for the accelerating scalar field in $\kappa$-deformed space-time, from the plane wave mode solutions of the $\kappa$-deformed Klein-Gordon equation. Further we compare this result with the one derived from the $\kappa$-deformed uncertainty relation and shows that in the small acceleration limit the former result reduces to the later one. 

This paper is organised in the following manner. In sec. 2, we introduce the definition of the $\kappa$-deformed space-time and other relevant details. Then we write down the Klein-Gordon equation in the $\kappa$-deformed space-time using the quadratic Casimir of the underlying symmetry algebra. In sec. 3, we obtain the plane wave solutions to the $\kappa$-deformed Klein-Gordon equation, valid up to first order in $a$ and calculate the power spectrum observed for this deformed mode solution, moving along a uniformly accelerating trajectory. From the expression for the $\kappa$-deformed power spectrum, we obtain the $\kappa$-deformed corrections to the Unruh temperature. In sec. 4, we calculate the $\kappa$-deformed corrections to Unruh temperature from the $\kappa$-deformed uncertainty relation and compare with the former one. Finally in sec. 5, we give the concluding remarks.

Here we use Minkowski metric with the signature $(-,+,+,+)$.

\section{$\kappa$-deformed Klein-Gordon equation}

In this section, we provide a brief description of the $\kappa$-deformed space-time and the underlying symmetry algebra associated with it. Using the quadratic Casimir of this symmetry algebra we construct Klein-Gordon equation in the $\kappa$-deformed space-time.

The $\kappa$-deformed space-time coordinates (i.e., $\hat{x}_{\mu}$) obey the following commutation relations \cite{kreal}
\begin{equation} \label{lie}
 [\hat{x}_i, \hat{x}_j]=0,~~[\hat{x}_0,\hat{x}_i]=ia\hat{x}_i,
\end{equation}
where $a$ is the $\kappa$-deformation length parameter. Note that we can obtain the corresponding commutative results in the limit $a\to 0$.

In this work, we follow the approach discussed in \cite{kreal,mel1,mel3}, where the $\kappa$-deformed space-times coordinates are represented as functions of the commutative coordinates and their derivatives. Thus we realise the $\kappa$-deformed space-time coordinates as
\begin{equation}\label{power1}
 \hat{x}_{i}=x_{i}\varphi(A),~~\hat{x}_0=x_0\psi(A)+iax_i\partial_i\gamma(A),
\end{equation} 
where $A=-ia\partial_0$. From Eq.(\ref{power1}) and Eq.(\ref{lie}), we get
\begin{equation}\label{power1a}
 \frac{\varphi'}{\varphi}\psi=\gamma-1,
\end{equation}
with the boundary conditions $\varphi(0)=1,~\psi(0)=1,~\gamma(0)=\varphi'(0)+1$ and $\varphi,~\psi,\gamma$ are finite. 

In \cite{kreal}, it has been shown that $\psi(A)=1$ and $\psi(A)=1+2A$ are the only two possible realisations of $\psi(A)$. So now onwards we work with the realisation $\psi(A)=1$. Next we choose $\varphi(A)=e^{-A}$ as this choice of $\varphi(A)$ is related to the $\kappa$-Poincare algebra in bi-cross product space \cite{majid}. Similarly we fix $\gamma(A)$ to be a constant. It is to be noted that these choices of realisations correspond to different ordering prescriptions \cite{kreal}.

The symmetry algebra of the $\kappa$-deformed space-time can be defined using the $\kappa$-Poincare algebra, where both the Poincare algebra as well as the explicit form of the generators are deformed, such that they reduce to their usual form in the commutative limit. Alternately the symmetry algebra of the $\kappa$-deformed space-time can also be realised in terms of the undeformed $\kappa$-Poincare algebra \cite{kreal}. The undeformed $\kappa$-Poincare algebra has exactly the same form as that of the usual Poincare algebra, but the explicit form of the generators are deformed. The undeformed $\kappa$-Poincare algebra is obtained to be of the following form \cite{kreal}
\begin{equation}\label{power4}
\begin{split}
 [\hat{M}_{\mu\nu},D_{\lambda}]&=\eta_{\nu\lambda}D_{\mu}-\eta_{\mu\lambda}D_{\nu},\\
 [D_{\mu},D_{\nu}]&=0,\\
 [\hat{M}_{\mu\nu},\hat{M}_{\lambda\rho}]&=\eta_{\mu\rho}\hat{M}_{\nu\lambda}+\eta_{\nu\lambda}\hat{M}_{\mu\rho}-\eta_{\nu\rho}\hat{M}_{\mu\lambda}-\eta_{\mu\lambda}\hat{M}_{\nu\rho}.
\end{split}
\end{equation}
The components of $\kappa$-deformed Lorentz generator is defined as \cite{kreal}
\begin{equation}\label{power5}
 \hat{M}_{\mu\nu}=\frac{\left(\hat{x}_{\mu}D_{\nu}-\hat{x}_{\nu}D_{\mu}\right)}{iaD_0+\sqrt{1+a^2D_{\alpha}D^{\alpha}}}.
\end{equation}
The derivative operator which transform as a four-vector under the undeformed $\kappa$-Poincare algebra is defined as Dirac derivatives $D_{\mu}$ \cite{kreal} and its components are defined as
\begin{equation}\label{power3}
D_{i}=\partial_{i},~~
D_{0}=\partial_{0}\frac{\sinh A}{A}-\frac{iae^{A}}{2}{\partial_{i} ^2},
\end{equation}
The quadratic Casimir corresponding to the undeformed $\kappa$-Poincare algebra is defined using Dirac derivative as \cite{kreal,mel4}
\begin{equation}\label{power6}
 D_{\mu}D^{\mu}=\Box\Big(1+\frac{a^2}{4}\Box\Big),
\end{equation}
where $\Box$ is the $\kappa$-deformed Laplacian and is defined as
\begin{equation}\label{power7}
 \Box=\partial_i^2e^A-\partial_0^2\frac{\sinh^2(A/2)}{(A/2)^2}
\end{equation}
We define the $\kappa$-deformed Klein-Gordon (KG) equation using the above defined quadratic Casimir, as \cite{mel4}
\begin{equation}\label{power10}
 \bigg(\Box\Big(1+\frac{a^2}{4}\Box\Big)-m^2\bigg)\phi(x)=0.
\end{equation}   
Here the above defined equations of motion (for the scalar field $\phi(x)$ in $\kappa$ space-time) is invariant under the undeformed $\kappa$-Poincare algebra. From this expression one can also write down the dispersion relation for the $\kappa$-deformed space-time (see \cite{mel4} for details).  


\section{$\kappa$-deformed Power spectrum}

In this section, we obtain a plane-wave solution to the $\kappa$-deformed KG equation, valid up to first order in $a$. Using this solution, we then calculate the power spectrum, corresponding to the outgoing mode of the massless $\kappa$-deformed scalar field, detected by an uniformly accelerating observer. From this we derive the expression for the modified Unruh temperature, valid up to first order in $a$. 

Here onwards we proceed the remaining analysis by considering the terms valid up to first order in $a$ only. Thus the $\kappa$-deformed KG equation (given in Eq.(\ref{power10}), valid up first order in $a$, becomes
\begin{equation}\label{power11}
 \Big(\partial^2_i-\partial^2_0-ia\partial_0\partial^2_i-m^2\Big)\phi(\vec{x},t)=0.
\end{equation}
We solve this equations of motion by choosing an ansatz for $\phi(\vec{x},t)$ as $\phi(\vec{x},t)=f(\vec{x})e^{-i\omega t}$. Substituting this in Eq.(\ref{power11}), the equations of motion reduces to the following form
\begin{equation}\label{power12}
 \Big(\partial^2_i-a\omega\partial^2_i+\omega^2-m^2\Big)f(\vec{x})=0.
\end{equation}
Now we multiply Eq.(\ref{power12}) throughout by $1+a\omega$ and keeping the terms valid up to first order in $a$, we get
\begin{equation}\label{power13}
 \Big(\partial^2_i+k_{eff}^2\Big)f(\vec{x})=0.
\end{equation}
We call the $\vec{k}_{eff}$ in the above expression as the effective wave number associated with $\kappa$-deformed space-time and this is defined as $\vec{k}_{eff}=(1+\frac{a\omega}{2})\vec{k}$ (where $|\vec{k}|=\sqrt{\omega^2-m^2}$). Hence from Eq.(\ref{power13}) we get $f(\vec{x})=e^{i\vec{k}_{eff}\cdot\vec{x}}$. Thus the plane-wave solution for the deformed KG equation (i.e., Eq.(\ref{power11})) is 
\begin{equation}\label{power14}
 \phi(\vec{x},t)=e^{-i(\omega t-\vec{k}_{eff}\cdot\vec{x})}.
\end{equation}
Note that here we have considered only the outgoing modes of the plane-wave solution. Using this general solution, we obtain the outgoing modes of the plane wave solution for a massless $\kappa$-deformed KG equation in the $(1+1)$ dimension as
\begin{equation}\label{power14a}
 \phi(x,t)=e^{-i\omega\big(t-x(1+\frac{a\omega}{2})\big)},
\end{equation}
where $\omega$ is the frequency and $k$ is the wave number associated with the plane wave propagating along the $x$ direction. 

Now we consider an uniformly accelerating frame which is known as the Rindler frame. The $(1+1)$ dimensional Minkowski space-time coordinates $(x,t)$ are related to the proper time $\tau$ of the Rindler observer and the constant acceleration $\mathcal{A}$ through the following coordinate transformation as
\begin{equation}\label{power15}
 x(\tau)=\frac{1}{\mathcal{A}}\cosh \mathcal{A}\tau,~t(\tau)=\frac{1}{\mathcal{A}}\sinh \mathcal{A}\tau.
\end{equation}
By substituting the parameterised form of the coordinates $x$ and $t$ (i.e., Eq.(\ref{power15})) in Eq.(\ref{power14a}), we get the explicit form of the $\kappa$-deformed outgoing mode, detected by the Rindler observer as
\begin{equation}\label{power16}
\begin{split}
 \phi\big(x(\tau),t(\tau)\big)
 =\exp{\bigg[\frac{i\omega}{\mathcal{A}}\Big(1+\frac{a\omega}{4}\Big)e^{-\mathcal{A}\tau}+\frac{ia\omega^2}{4\mathcal{A}}e^{\mathcal{A}\tau}\bigg]}.
\end{split}
\end{equation}
In order to obtain the power spectrum of the outgoing mode, as detected by the accelerating observer, we first calculate the Fourier transform of this mode (with respect to $\tau$) using the definition $g(\nu)=\int d\tau\phi(\tau)e^{-i\nu\tau}$. The $\nu$ in this above definition corresponds to an arbitrarily chosen frequency from the continuous spectrum of frequencies available to the accelerating observer. 

Therefore by substituting the explicit form of the deformed outgoing mode (i.e., Eq.(\ref{power16})) in the definition $g(\nu)=\int d\tau\phi(\tau)e^{-i\nu\tau}$, we get
\begin{equation}\label{power18}
 g(\nu)=\int_{-\infty}^{\infty} d\tau\exp{\bigg[\frac{i\omega}{\mathcal{A}}\Big(1+\frac{a\omega}{4}\Big)e^{-\mathcal{A}\tau}+\frac{ia\omega^2}{4\mathcal{A}}e^{\mathcal{A}\tau}-i\nu\tau\bigg]}
\end{equation}
We simplify the above integral by substituting $u=e^{-\mathcal{A}\tau}$. Thus we have
\begin{equation}\label{power19}
 g(\nu)=\frac{1}{\mathcal{A}}\int_0^{\infty}du~e^{\frac{i\omega}{\mathcal{A}}(1+\frac{a\omega}{4})u}~u^{(\frac{i\nu}{\mathcal{A}}-1)}~e^{\frac{ia\omega^2}{4\mathcal{A}}\frac{1}{u}}
\end{equation}
From the above expression, we see that $e^{\frac{ia\omega^2}{4\mathcal{A}}\frac{1}{u}}$ term is completely contributed due to the $\kappa$-deformation of the plane wave solution. So we expand this $e^{\frac{ia\omega^2}{4\mathcal{A}}\frac{1}{u}}$ and keep the terms valid up to first order in $a$. Thus Eq.(\ref{power19}) becomes
\begin{equation}\label{power20}
 g(\nu)=\frac{1}{\mathcal{A}}\int_0^{\infty}du~e^{\frac{i\omega}{\mathcal{A}}(1+\frac{a\omega}{4})u}~u^{(\frac{i\nu}{\mathcal{A}}-1)}+\frac{ia\omega^2}{4\mathcal{A}^2}\int_0^{\infty}du~e^{\frac{i\omega}{\mathcal{A}}(1+\frac{a\omega}{4})u}~u^{(\frac{i\nu}{\mathcal{A}}-2)}
\end{equation}
We evaluate the above integrals using the relation $\int^{\infty}_0x^{s-1}e^{-qx}dx=\Gamma(s)e^{-s\ln q}$ \cite{table}. This gives the explicit form of the Fourier transform associated with the deformed mode as
\begin{equation}\label{power21}
 g(\nu)=\frac{1}{\mathcal{A}}\bigg[\frac{\omega}{\mathcal{A}}\Big(1+\frac{a\omega}{4}\Big)\bigg]^{-\frac{i\nu}{\mathcal{A}}}\exp{\big(-\pi\nu/2\mathcal{A}\big)}\Gamma(i\nu/\mathcal{A})\bigg(1-\frac{a\omega^3}{4\mathcal{A}^2\big(1-i\nu/\mathcal{A}\big)}\bigg)
\end{equation}
Next we calculate the power spectrum per logarithmic band in frequency using the definition $P(\nu)=\nu|g(\nu)|^2$ \cite{paddy,paddy1,hammad}. By taking the square of the modulus of the Fourier transform (given in Eq.(\ref{power21})) and substituting it in this definition, we obtain the power spectrum detected by Rindler observer in $\kappa$-deformed space-time as
\begin{equation}\label{power23}
 P(\nu)=\frac{2\pi}{\mathcal{A}}\Big(e^{2\pi\nu/\mathcal{A}}-1\Big)^{-1}\bigg[1-\frac{a\omega^3}{2\mathcal{A}^2\big(1+\nu^2/\mathcal{A}^2\big)}\bigg].
\end{equation}
We observe that the power spectrum (associated with the outgoing modes of the $\kappa$-deformed KG field) depends on the frequency $\omega$ and thus on the energy of the particle. This energy dependent correction term in the power spectrum is induced by the modified dispersion relation associated with the space-time non-commutativity. Due to this correction term, the Rindler observer will detect a reduced power spectrum as compared to its commutative counterpart.

The first order correction term of the power spectrum obtained in \cite{glik} is found to be have linear dependence on the acceleration compared to our result shown in Eq.(\ref{power23}). This difference is expected to arise due to the ambiguity in the ordering of star product.

As in the commutative case the power spectrum in the $\kappa$-deformed space-time retains the crucial $(e^{2\pi\nu/\mathcal{A}}-1)^{-1}$ factor, representing the Planckian spectrum. But on the other hand we find that this spectrum is modified due to the $(1-\frac{a\omega^3}{2\mathcal{A}^2(1+\nu^2/\mathcal{A}^2)})$ factor. This modification is expected to deform the temperature associated with this power spectrum. Now let us calculate this deformed temperature. For this purpose we identify this modification factor as $(1-\beta)$, where $\beta=\frac{a\omega^3}{2 \mathcal{A}^2(1+\nu^2/\mathcal{A}^2)}$. Since $\beta<<1$ we take $1-\beta\simeq e^{-\beta}$. Using this in Eq.(\ref{power23}), we get
\begin{equation}\label{power24}
 P(\nu)\simeq\frac{2\pi}{\mathcal{A}}\Big(e^{\frac{2\pi\nu}{\mathcal{A}}+\beta}-1\Big)^{-1}.
\end{equation}
Note that we have neglected the term linear in $\beta$ in the denominator of the RHS of Eq.(\ref{power24}) as in \cite{scar1}. Now we compare the power spectrum obtained in Eq.(\ref{power24}) with the standard Planckian distribution at a temperature $T$, i.e., $P(\nu)=\frac{2\pi}{\mathcal{A}}(e^{\nu/T}-1)^{-1}$ \cite{unruh}. This gives the explicit form of the deformed temperature as 
\begin{equation}\label{power25}
 T=T_U\bigg(1-\frac{a\omega^3}{4\pi\nu \mathcal{A}\big(1+\nu^2/\mathcal{A}^2\big)}\bigg),
\end{equation}
where $T_U=\frac{\mathcal{A}}{2\pi}$ is the Unruh temperature \cite{unruh}. Thus we find that the Unruh temperature picks up an energy dependent correction term under the $\kappa$-deformation. We notice that the effect of this non-commutativity is to decrease the value of the usual Unruh temperature. Similar results were obtained while calculating the modified Unruh temperature associated with the non-local theories \cite{gim}. 

Note that $\omega$ is the frequency defined by the Minkowski observer which is distributed over the continuous range of frequencies $\nu$ defined by the Rindler observer, such that it peaks at the value $\nu=\omega$ \cite{paul}. Therefore it is reasonable to take $\omega=\nu$ and thus the expression for the deformed Unruh temperature becomes
\begin{equation}\label{power25z}
 T=T_U\bigg(1-\frac{a\nu^2}{4\pi \mathcal{A}\big(1+\nu^2/\mathcal{A}^2\big)}\bigg),
\end{equation}
The behaviour of the Unruh temperature in $\kappa$-deformed space-time can be understood in detail by considering the following cases.

\textbf{Case 1: Small acceleration limit}

In the small acceleration limit, we assume the acceleration is much less than the detectable frequency, i.e., $\mathcal{A}<<\nu$. Thus the factor $\big(1+\nu^2/\mathcal{A}^2\big)$ in Eq.(\ref{power25z}) can be taken as $1+\frac{\nu^2}{\mathcal{A}^2}\simeq\frac{\nu^2}{\mathcal{A}^2}$ and therefore Eq.(\ref{power25z}) reduces to the following form 
\begin{equation}\label{power25a}
 T=T_U\bigg(1-\frac{aT_U}{2}\bigg).
\end{equation}  
The above expression (i.e., Eq.(\ref{power25a}) does not depends of the energy, unlike the expression obtained in Eq.(\ref{power25z}). Thus we observe that in the small acceleration limit, the Unruh temperature in $\kappa$-deformed space-time becomes energy independent. We also find that the first order correction is quadratic in the Unruh temperature.

\textbf{Case 2: Large acceleration limit}

In the large acceleration limit, we assume the acceleration is much large compared to the frequency, i.e., $\mathcal{A}>>\nu$. Thus the factor $\big(1+\nu^2/\mathcal{A}^2\big)^{-1}$ can be approximated as $\big(1+\nu^2/\mathcal{A}^2\big)^{-1}\simeq 1-\frac{\nu^2}{\mathcal{A}^2}\simeq 1$. Hence Eq.(\ref{power25z}) becomes
\begin{equation}\label{power25y}
 T=T_U\bigg(1-\frac{a\nu^2}{8\pi^2 T_U}\bigg).
\end{equation}
We find that in the large acceleration limit the $a$ dependent term does not contain the $T_U$ term. Thus compared to the small acceleration limit, here the first order correction term depends only on the energy of the particle.

\section{Unruh temperature from $\kappa$-deformed uncertainty principle}

In this section we provide a heuristic derivation of the Unruh temperature from $\kappa$-deformed uncertainty relation, by following the arguments given in \cite{scar,scar1,wschung,wschung1}. We then show that this expression is in exact agreement with the expression for the deformed Unruh temperature obtained from the small acceleration limit of the $\kappa$-deformed power spectrum.

We consider a particle moving with a uniform acceleration $\mathcal{A}$. This particle acquires kinetic energy as it accelerates along a distance $\delta x$ and the resulting energy creates $N$ particle pairs due to the vacuum fluctuations. Thus we have
\begin{equation}\label{power25x}
 m\mathcal{A}\delta x\simeq 2Nm.
\end{equation}
We assume this $N$ particle pair created is localised within a region of width $\delta x$, given by
\begin{equation}\label{power26}
 \delta x\simeq 2N/\mathcal{A}.
\end{equation}
Now let us use the $\kappa$-deformed uncertainty relation between the position and momenta coordinate obtained from \cite{anjana}, i.e.,
\begin{equation}\label{power29}
 \Delta x\Delta p\Big(1+\frac{a}{2\Delta x}\Big)\geq\frac{1}{2}
\end{equation}
Thus the energy fluctuation of the particles, localised within a region of width $\delta x$, in the $\kappa$-deformed space-time can be obtained by taking $\Delta p=\Delta E$ in the $\kappa$-deformed uncertainty relation, given in Eq.(\ref{power29}). Hence we get the energy fluctuations in the $\kappa$-deformed space-time to be of the following form
\begin{equation}\label{power30}
 \delta E\simeq\frac{1}{2\delta x}-\frac{a}{4(\delta x)^2}.
\end{equation}
Substituting Eq.(\ref{power26}) in the above relation, we get the fluctuations in the energy as
\begin{equation}\label{power30a}
 \delta E\simeq\frac{\mathcal{A}}{4N}-\frac{a\mathcal{A}^2}{16N^2}
\end{equation}
We assume that this energy fluctuation is caused due to the thermal energy, i.e., $\delta E=\frac{1}{2}T$, of the particle. Using this in the above relation and choosing $N=\pi$ (i.e., $N\simeq 3$ particle pairs), we obtain the expression for the modified temperature as
\begin{equation}\label{power31}
 T=T_U\Big(1-\frac{aT_U}{2}\Big).
\end{equation}
From the above (i.e., Eq.(\ref{power31}), we find that the deformed Unruh temperature obtained from the $\kappa$-deformed uncertainty principle is exactly equal to that obtained from the $\kappa$-deformed power spectrum, in the small acceleration limit (i.e., Eq.(\ref{power25y})).

Recently the Unruh temperature has been measured experimentally to be $T=(1.80\pm 0.51)$ PeV \cite{lynch}. By comparing the error bar of this result with the first order correction term of the deformed Unruh temperature given in Eq.(\ref{power31}), we obtain an upper bound on the $\kappa$-deformation parameter as $a\leq 10^{-23}$m.

\begin{figure}[!htb]\centering
\includegraphics[width=0.65\textwidth]{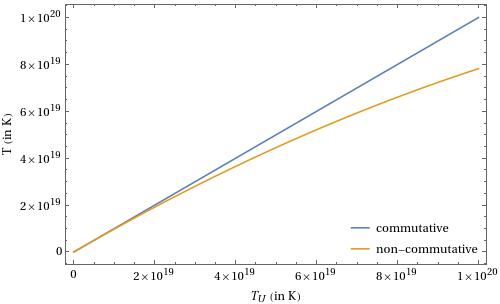}
\caption{Variation of $T$ against $T_U$ (for $a\sim 10^{-23}$m)}
\label{fig:unrh}
\end{figure}

In Fig.(\ref{fig:unrh}) we analyse the behaviour of $\kappa$ modified Unruh temperature (given in Eq.(\ref{power31})) with respect to the commutative Unruh temperature and we observe a pronounced deviation of $T$ from $T_U$ at higher temperature regimes. Such an apparent deviation is caused due to the minimal length scale effects associated with the non-commutativity of space-time.


\section{Conclusion}

We have studied the emission power spectrum due to Unruh effect in the $\kappa$-deformed space-time, by calculating the power spectrum of an outgoing scalar field mode (obeying the $\kappa$-deformed KG equation) detected by an uniformly accelerating observer. We see that due to the non-commutativity, the value of the power spectrum in the $\kappa$-deformed space-time reduces in comparison with the commutative space-time. The first order correction term of the power spectrum is shown to depend on the energy of the particle, as a consequence of the modified dispersion relation arising from the space-time non-commutativity. The $\kappa$-deformed corrections to the power spectrum obtained using the star product method has also been shown to have an energy dependence \cite{glik}. The energy dependent correction term in the power spectrum of the uniformly accelerating scalar field compatible with the generalised uncertainty principle (GUP) has been reported in \cite{ecv,sg}. Those energy dependence has been shown to arise from the modified dispersion relation associated with the GUP models. 

The emission power spectrum associated with the $\kappa$-deformed KG field is found to have a Planckian distribution, with a modified temperature and this gives the expression for the Unruh temperature in the $\kappa$-deformed space-time. The $\kappa$-deformed Unruh temperature is shown to have an energy dependent correction term. Similar energy dependent corrections to the Unruh temperature has been shown to arise in the non-local field theories \cite{gim}. Thus we observe that the Unruh temperature, arising from the theories with a non-local nature or with a modified dispersion relation, is expected to have an energy dependent correction term. In our result, we find that in the small acceleration limit, the first order correction term is independent of the energy and depends quadratically on the acceleration. But the situation is different in the large acceleration limit where the first order correction term does not depend on the acceleration and it depends quadratically on the energy.

We heuristically derive the expression for the Unruh temperature in the $\kappa$-deformed space-time from the $\kappa$-deformed uncertainty principle. We find that the expression for $\kappa$-deformed Unruh temperature obtained from the $\kappa$-deformed uncertainty principle is exactly equal to that derived from the small acceleration limit of the $\kappa$-deformed power spectrum method. We also observe a noticeable deviation of the modified Unruh temperature from its commutative version at high energy scales. Similar deviation for the modified Unruh temeprature in GUP formalism at high energy scale has also been shown in \cite{wschung}. Finally we obtain an upper bound on the $\kappa$-deformation parameter as $a\leq10^{-23}$m, by comparing the deformed Unruh temperature with the experimental result. This bound has also been reported in \cite{suman} while deriving the $\kappa$-deformed corrections to Casimir energy. 


\subsection*{\bf Acknowledgements}

The author would like to acknowledge the Vice-Chancellor, Prof. Rajendrakumar Anayath for the support. The author also thank Prof. E. Harikumar for useful suggestions. 

\subsection*{\bf Data availability}

Data sharing is not applicable to this article as no datasets were generated or analysed during the current study.


\begin{thebibliography}{99}

\bibitem{snyder}
H. S. Snyder, \textit{Phys. Rev. }\textbf{71}, 38 (1947).
\bibitem{connes} 
A. Connes, \textit{Noncommutative Geometry }(Academic Press, London, 1994).
\bibitem{doplicher}
S. Doplicher, K. Fredenhagen and J. E. Roberts, \textit{Phys. Lett. }\textbf{B 331} (1994) 29; \textit{Commun. Math. Phys. }\textbf{172}, 187 (1995).
\bibitem{sieberg}
N. Seiberg and E. Witten, \textit{JHEP} \textbf{032}, 9909 (1999).
\bibitem{chai}
M. Chaichian, P. P. Kulish, K. Nishijima and A. Tureanu, \textit{Phys. Lett. }\textbf{B 604}, 98 (2004); M. Chaichian, P. Presnajder and A. Tureanu, \textit{Phys. Rev. Lett. }\textbf{94}, 151602 (2005).
\bibitem{loop}
F. Cianfrani, J. Kowalski-Glikman, D. Pranzetti and G. Rosati, \textit{Phys. Rev. }\textbf{D 94}, 084044 (2016).
\bibitem{wjk}
J. Lukierski, A. Nowicki, and H. Ruegg, \textit{Phys. Lett. } \textbf{B 293}, 344 (1992); J. Lukierski and H. Ruegg, \textit{Phys. Lett. }\textbf{B 329}, 189 (1994); J. Lukierski, H. Ruegg and W. J. Zakrewski, \textit{Ann. Phys.} \textbf{243}, 90 (1995).
\bibitem{majid}
S. Majid and H. Ruegg, \textit{Phys. Lett. }\textbf{B 334}, 348 (1994).
\bibitem{gauge0}
M. Dimitrijevi, L. Jonke, L. Moller, E. Tsouchnika, J. Wess and M. Wohlgenann, \textit{Eur. Phys. J. }\textbf{C 31}, 129 (2003);
M. Dimitrijevic, F. Meyer, L. Moller and J. Wess, \textit{Eur. Phys. J. }\textbf{C 36}, 117 (2004); M. Dimitrijevic, L. Jonke and L. Moller, \textit{JHEP} \textbf{086}, 0509 (2005); M. Dimitrijevic and L. Jonke, \textit{JHEP} \textbf{12}, 080 (2011); M. Dimitrijevic, L. Jonke and A. Pachol, \textit{SIGMA} \textbf{10}, 063 (2014).
\bibitem{kreal}
S. Meljanac and M. Stojic, \textit{Eur. Phys. J. }\textbf{C 47}, 531 (2006).
\bibitem{mel1}
S. Meljanac, S. Kresic-Juric and M. Stojic, \textit{Eur. Phys. J. }\textbf{C 51}, 229 (2007).
\bibitem{mel3}
S. Meljanac, A. Samsarov, M. Stojic, and K. S. Gupta, \textit{Eur. Phys. J. }\textbf{C 53}, 295 (2008).
\bibitem{mel4}
T. R. Govindarajan, K. S. Gupta, E. Harikumar, S. Meljanac and D. Meljanac, \textit{Phys. Rev. }\textbf{D 80}, 025014 (2009).
\bibitem{mel5}
S. Meljanac, A. Samsarov, J. Trampetic and M. Wohlgenannt, \textit{JHEP} \textbf{12}, 010 (2011).
\bibitem{unruh}
W. G. Unruh, \textit{Phys. Rev. }\textbf{D 14}, 870 (1976).
\bibitem{sachin}
N. Acharyya and S. Vaidya, \textit{JHEP} \textbf{09}, 045 (2010).
\bibitem{vishnu2}
E. Harikumar and V. Rajagopal, \textit{Nucl. Phys. }\textbf{B 974}, 115633 (2022).
\bibitem{yee}
H-C. Kim, J. H. Yee and C. Rim, \textit{Phys. Rev. }\textbf{D 75}, 045017 (2007).
\bibitem{verma1}
E. Harikumar, A. K. Kapoor and R. Verma, \textit{Phys. Rev. }\textbf{D 86}, 045022 (2012).
\bibitem{verma2}
E. Harikumar and R. Verma, \textit{Mod. Phys. Lett. }\textbf{A 28}, 1350063 (2013).
\bibitem{vishnu1}
E. Harikumar and V. Rajagopal, \textit{Eur. Phys. J. }\textbf{C 79}, 735 (2019).
\bibitem{glik}
J. Kowalski-Glikman, \textit{Phys. Rev. }\textbf{D 8}, 044028 (2009).
\bibitem{table}
I. S. Gradshteyn and I. M. Ryzhik, \textit{Table of Integrals, Series and Products} (Academic Press, 2007).
\bibitem{paddy}
K. Srinivasan, L. Sriramkumar and T. Padmanabhan, \textit{Phys. Rev. }\textbf{D 56}, 6692 (1997).
\bibitem{paddy1}
T. Padmanabhan, \textit{Gravitation, Foundations and Frontiers} (Cambridge University Press, Cambridge, UK, 2010).
\bibitem{hammad}
F. Hammad, A. Landry and D. Dijamco, \textit{Phys. Rev. }\textbf{D 103}, 085010 (2021).
\bibitem{scar1} 
F. Scardigli, M. Blasone, G. Luciano and R. Casadio, \textit{Eur. Phys. J. }\textbf{C 78}, 728 (2018).
\bibitem{gim} 
Y. Gim, H. Um and W. Kim, \textit{Phys. Lett. }\textbf{B 784}, 206 (2018).
\bibitem{paul}
P. M. Alsing and P. W. Milonni, \textit{Am. J. Phys. }\textbf{72}, 1524 (2004).
\bibitem{scar} 
F. Scardigli, \textit{Nuovo Cim. }\textbf{B 110}, 1029 (1995).
\bibitem{wschung}
W. S. Chung and H. Hassanabadi, \textit{Phys. Lett. }\textbf{B 793}, 451 (2019).
\bibitem{wschung1}
H. Hassanabadi, N. Farahani, W. S. Chung and B. C. Lutfuoglu, \textit{Eur. Phys. Lett. }\textbf{130}, 40001 (2020).
\bibitem{anjana}
V. Anjana, E. Harikumar and A. K. Kapoor, \textit{Int. J. Mod. Phys. }\textbf{A 32}, 1750183 (2017).
\bibitem{lynch} 
H. M. Lynch, E. Cohen, Y. Hadad and I. Kaminer, \textit{Phys. Rev. }\textbf{D 104}, 025015 (2021).
\bibitem{ecv} 
B. R. Majhi and E. C. Vagenas, \textit{Phys. Lett. }\textbf{B 725}, 477 (2013).
\bibitem{sg} 
A. Mukherjee, S. Gangopadhyay and M. Dutta, \textit{Eur. Phys. Lett.} \textbf{129}, 30002 (2020). 
\bibitem{suman}
E. Harikumar, S. K. Panja and V. Rajagopal, \textit{Nucl. Phys. }\textbf{B 950}, 114842 (2020).

\end{thebibliography}
\end{document}